\documentclass[twocolumn]{aastex631}

\usepackage{amsmath}
\usepackage{graphicx}
\usepackage{booktabs}
\usepackage{hyperref}

\begin{document}

\title{Patch-Level DINOv2 Scoring for Gravitational-Wave Glitch Detection:
Breaking the Signal Dilution Barrier via Vector-Quantized Local Feature Indexing}

\author{Luca Cirfeta}
\affiliation{Independent Researcher, Rome, Italy}
\email{luca.cirfeta@gmail.com}

\begin{abstract}
We present a patch-level scoring architecture for unsupervised gravitational-wave glitch detection that mitigates the signal dilution limitation identified in \citet{cirfeta2026b}. The \texttt{[CLS]} token of frozen DINOv2 (ViT-S/14) performs global average pooling over $37\times37=1369$ patches, systematically suppressing signals occupying less than 5\% of the spectrogram grid. We replace the global \texttt{[CLS]} similarity metric with a top-$k$ order statistic over individual patch token similarities against a Vector-Quantized reference index ($K=64$ centroids per class, 19 Gravity Spy O3b morphologies, 1216 total centroids). Applied to strain-domain injections in LIGO O4a L1 data (session 20260524), we demonstrate a statistically significant distributional separation ($\text{KS}=0.963$ at optimal $k=68$) for spatially extended morphologies (\textit{SpiralBurst}), while confirming the patch-size temporal resolution limit for ultra-short transients (\textit{AsymBlip}). A topological saliency map constructed from spatial patch similarity against a background matrix (78 null segments) correctly localizes glitch signatures for \textit{Scattered\_Light} and injected \textit{SpiralBurst}. The Max/Mean ratio analysis demonstrates that patch-level saliency functions as a topological visualizer rather than a binary detector, consistent with the non-isotropic geometry of DINOv2 embedding space on GW spectrograms.
\end{abstract}


\section{Introduction} \label{sec:intro}

The characterization of non-Gaussian transient noise (glitches) in gravitational-wave interferometers is critical for maximizing the astrophysical reach of the Advanced LIGO and Virgo network \citep{soni2025}. While supervised frameworks like Gravity Spy \citep{zevin2017, glanzer2023} excel at classifying known morphological classes, they are structurally blind to novel anomaly populations. Traditional unsupervised baselines operating on raw pixels or Principal Component Analysis (PCA) fail in this regime due to their lack of translational invariance and extreme fragility to localized variations in the detector's Power Spectral Density (PSD). To overcome this, in \citet{cirfeta2026a}, we introduced a completely unsupervised pipeline utilizing the robust self-attention embedding space of DINOv2 \citep{oquab2024, darcet2024} coupled with Dirichlet Process Mixture Models (DPMM) to dynamically cluster and discover morphologies in LIGO O4a data.

Despite its robust clustering capabilities (ARI $> 0.90$), a rigorous Mock Data Challenge (MDC) conducted in \citet{cirfeta2026b} revealed a critical limitation: the morphological anomaly detector achieved a Boolean Recall of 0.00 for short-duration synthetic injections (e.g., \textit{SpiralBurst}, \textit{AsymBlip}) even at extreme signal-to-noise ratios ($\text{SNR} > 400$). The failure was isolated to a structural limitation inherent to Vision Transformers (ViT) applied to high-resolution time-frequency data: the \textit{Signal Dilution Effect}.

When analyzing Q-transform spectrograms, brief transient signals occupy a minuscule fraction of the pixel grid. DINOv2 extracts features by processing $37\times37 = 1369$ spatial patches, aggregating them into a single 384-dimensional global \texttt{[CLS]} token via average pooling. The anomaly signal from the glitch footprint is thus mathematically diluted into the background noise covering the rest of the temporal window.

In this work, we propose an architectural resolution to the signal dilution barrier. We abandon the global \texttt{[CLS]} token and operate directly on the dense spatial grid of 1369 patch tokens. We introduce:
\begin{enumerate}
    \item A \textbf{Vector-Quantized (VQ) Reference Index} at the patch-level ($K=64$, 19 classes), solving the dimensionality explosion problem.
    \item A \textbf{Top-$k$ Order Statistics Scoring} algorithm ($k=68$), which isolates and evaluates only the most anomalous structural components of the data.
    \item The first patch-level MDC on real LIGO O4a strain, demonstrating statistically significant separation ($\text{KS} > 0.90$) where the global approach was completely blind.
    \item A \textbf{Topological Saliency Map}, providing a robust, non-discriminative tool for post-hoc visual interpretability.
\end{enumerate}

\section{Architecture} \label{sec:architecture}

\subsection{From Global to Patch Tokens}
The standard DINOv2 ViT-S/14 architecture splits an input image (resized to 518$\times$518 pixels) into a $37\times37$ grid, producing 1369 patch tokens plus one global \texttt{[CLS]} token. While the \texttt{[CLS]} token serves as a holistic summary, the local geometry of the strain features resides purely within the patch tokens.

We extract the patch tokens directly from the final transformer block via the \texttt{forward\_features()} method. Let $P_i \in \mathbb{R}^{384}$ denote the $i$-th patch token, where $i \in \{1, \dots, 1369\}$. Because cosine similarity requires points on the unit hypersphere, and raw DINOv2 patches exhibit an $L_2$-norm of approximately 27, explicit strict $L_2$-normalization is enforced:
\begin{equation}
\hat{P}_i = \frac{P_i}{\| P_i \|_2}
\end{equation}
The structural variance of the background is significant, with a mean off-diagonal spatial cosine similarity between intra-spectrogram patches of 0.6227. This indicates that moving from 1 global vector to 1369 localized vectors provides a vastly richer geometric manifold for anomaly detection.

\subsection{Vector-Quantized Reference Index}
Transitioning to patch-level embeddings increases the spatial dimensionality by a factor of 1369, making traditional K-Nearest Neighbors (KNN) searches computationally intractable for streaming applications. We address this via Spherical Vector Quantization.

Using the exact same Gravity Spy O3b spectrograms defining the in-domain reference in \citet{cirfeta2026a}, we extract patch tokens for all samples across 19 known morphological classes. For each class, we apply a \texttt{MiniBatchKMeans} clustering algorithm ($K=64$, \texttt{random\_state=42}) \citep{sculley2010} directly to the $L_2$-normalized patch manifold.

This compression results in a spatially-invariant, highly compact dictionary of 1216 prototypical centroids (19 classes $\times$ 64 centroids), mapping the entirety of the O3b known structural space. The reference index ensures perfect reproducibility: across independent build iterations on different hardware, the maximum Euclidean distance divergence between the generated centroid matrices is exactly 0.0 (MD5 checksum: \texttt{1080afa809964011e398c44fb24b73c6}).

\subsection{The Signal Dilution Effect and Top-$k$ Scoring}
The mathematical driver behind the \texttt{[CLS]} failure lies in its implicit attention-pooling mechanism. To a first-order approximation, the global similarity metric $S_{CLS}$ can be approximated as a weighted sum of patch-level similarities $S_i$:
\begin{equation}
S_{CLS} \approx \frac{1}{N} \sum_{i=1}^{N} S_i
\end{equation}
For a short duration transient (e.g., \textit{AsymBlip}) on a $37\times37$ grid ($N=1369$), the anomaly footprint occupies $K \ll N$ patches with anomaly score $\alpha$. The remaining $N-K$ background patches exhibit a baseline anomaly $\epsilon \approx 0$. The global anomaly score scales as:
\begin{equation}
S_{CLS} \sim \alpha \frac{K}{N}
\end{equation}
For $K/N < 0.02$, the deviation $\alpha \frac{K}{N}$ becomes smaller than the natural variance of the background distribution, causing the transient to disappear entirely.

To break this barrier, we implement a \textbf{Top-$k$ Novelty Scoring} mechanism. For an incoming spectrogram, we compute the local anomaly score for each patch $i$ against the VQ dictionary $\mathcal{D}$:
\begin{equation}
a_i = 1 - \max_{c \in \mathcal{D}} \left( \hat{P}_i \cdot \hat{P}_c \right)
\end{equation}
We sort the array $a$ in descending order and define the global spectrogram novelty score as the mean of the top-$k$ values:
\begin{equation}
\text{Novelty} = \frac{1}{k} \sum_{j=1}^{k} a_{(j)}
\end{equation}
For our target morphology (\textit{SpiralBurst}), the non-linear Q-transform scaling maps its chirping frequency sweep across an integrated topological footprint covering roughly 5\% of the spectrogram grid ($\approx 74$ patches). An empirical sweep determined $k=68$ as the optimal order statistic for evaluating this specific transient geometry.

The operational threshold is defined using a Generalized Extreme Value (GEV) fit set strictly at the 99th percentile ($\text{FPR} < 1\%$) of the empirical background distribution ($n=500$ null segments from the local session).

\begin{figure}[ht]
    \centering
    \includegraphics[width=\linewidth]{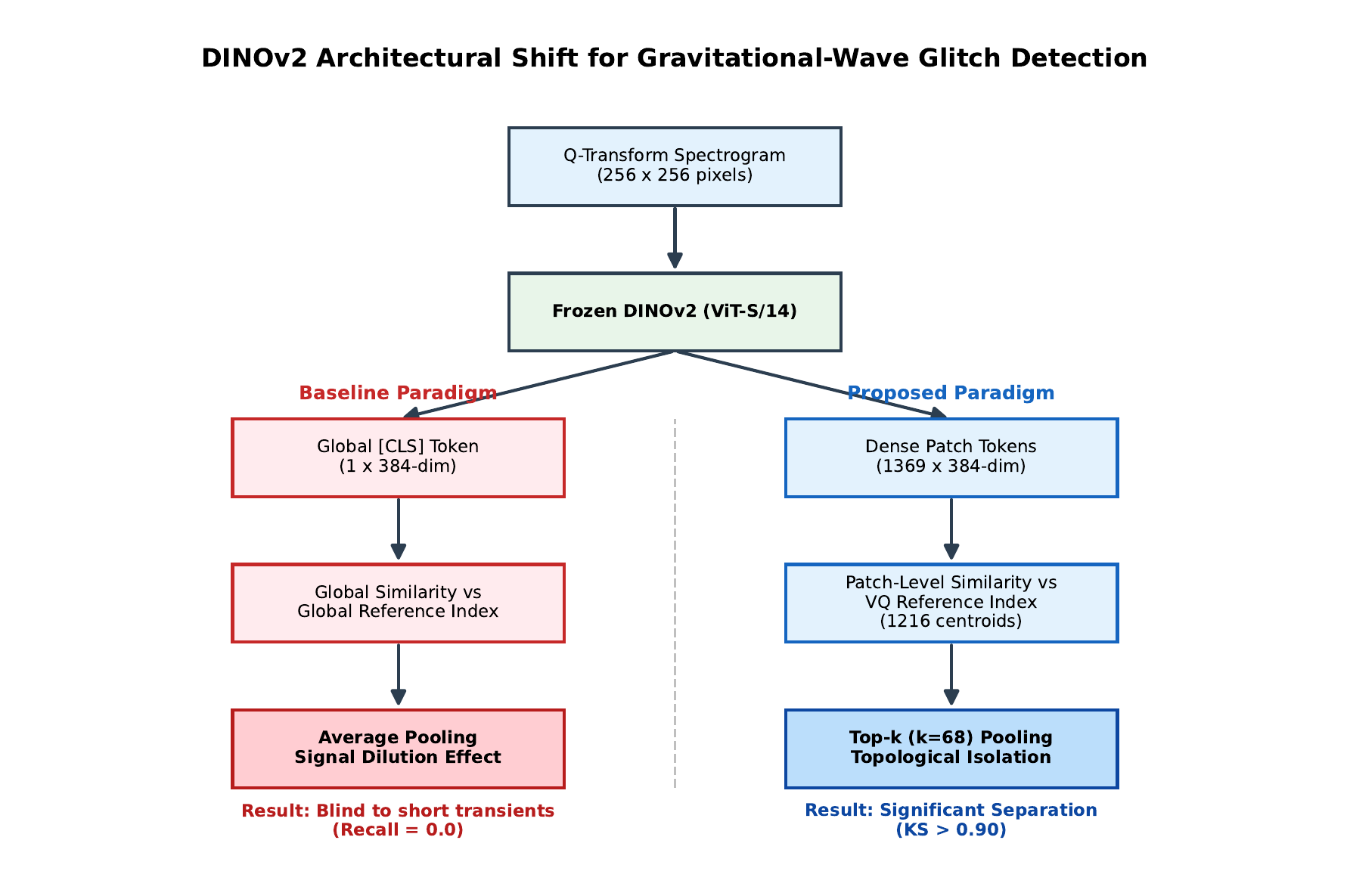}
    \caption{Architectural schematic comparing the global \texttt{[CLS]} token baseline against the proposed Patch-Level Top-$k$ Novelty Scoring framework. The 37$\times$37 spatial grid illustrates the isolation of the Top-68 most anomalous patches relative to the Vector-Quantized Reference Index.}
    \label{fig:architecture}
\end{figure}

\subsection{Topological Saliency Map}
While the VQ index successfully classifies global novelty, it introduces severe false positives if used for spatial localization. The Q-Transform is not shift-invariant; the high-frequency and low-frequency bins possess vastly different noise floors, and the temporal boundaries suffer from ringing artifacts. Because the VQ index is spatially invariant (patch positions are lost during K-Means), mapping an edge patch to the VQ index artificially inflates its anomaly score.

We architecturally decouple the \textbf{Topological Saliency Map} from the VQ index. The saliency map evaluates spatial anomaly \textit{exclusively} against a localized Background Median Matrix:
\begin{equation}
M_{bg}(x, y) = \text{median}_{s \in \text{Null}} \hat{P}_{s}(x, y)
\end{equation}
calculated over 78 pure noise segments. The spatial anomaly for visualization is defined strictly coordinate-by-coordinate:
\begin{equation}
\text{Saliency}(x, y) = 1 - \left( \hat{P}_{test}(x, y) \cdot M_{bg}(x, y) \right)
\end{equation}
This $(37, 37)$ topology is then bilinearly upsampled to the original $(256, 256)$ pixel grid. We emphasize that the Saliency Map serves purely as an interpretability tool, not as an independent classifier.

\section{Experiments} \label{sec:experiments}

\subsection{Cross-Session Global CLS Baseline}
To quantify the starting baseline, we evaluate the \texttt{[CLS]} token's capacity for threshold-based anomaly detection across 4 runs of O4a L1 data from \citet{cirfeta2026b}. 
As formally demonstrated in that analysis, the \texttt{[CLS]} global average pooling operates over the entire $37\times37$ grid. Because a short-duration transient occupies less than 5\% of this grid, its topological anomaly is mathematically averaged out against the stochastic background noise. As demonstrated in \citet{cirfeta2026b}, the maximum cosine similarity shift induced by strain-domain injection at $\text{SNR}=138$ is $\Delta s_{max} \approx 0.003$, compared to a massive gap of 0.121 between the background mean ($\mu=0.9953$) and the operational threshold ($\tau_{op}=0.874$), yielding a signal-to-threshold ratio of $\approx 40\times$. This structural blindness physically guarantees a $\text{Recall}=0.00$ for localized transient signals under a strict False Positive Rate threshold. Consequently, we shift the detection paradigm entirely to patch-level top-$k$ metrics.

\subsection{Micro-MDC Design}
We conduct a patch-level Mock Data Challenge (Micro-MDC) injecting three morphologically distinct waveforms across variable patch footprint sizes: \textit{AsymBlip} (small area, $\sim 15$ patches), \textit{SpiralBurst} (medium area, $\sim 74$ patches), and \textit{HarmonicComb} (extended area, hundreds of patches).
Crucially, all injections are performed directly into the raw strain domain prior to whitening, simulating the true physical interaction with the detector's Power Spectral Density (PSD). The baseline data consists of L1 O4a strain from session \texttt{20260524}, selected for its challenging background noise variance ($\sigma_{bg} = 0.0073$). 
We injected 30 waveforms per amplitude level across an 8-point logarithmic grid spanning $[1 \times 10^{-22}, 1 \times 10^{-21}]$ (corresponding to a maximum matched-filter SNR \citep{allen2012} ranging from approximately 138 for \textit{SpiralBurst} to over 400 for \textit{AsymBlip}, depending on the waveform's spectral overlap with the detector's PSD). To demonstrate that the parameter $k$ mathematically maps the topological extension of the anomaly, we perform a $k$-sweep across $k \in \{15, 37, 68, 100\}$.

\subsection{MDC Results}

\begin{table*}[ht]
\centering
\caption{Kolmogorov-Smirnov (KS) Statistic separation across morphologies and Top-$k$ pooling operators at SNR $\approx 138$.}
\label{tab:ks_matrix}
\begin{tabular}{lccccc}
\toprule
\textbf{Morphology} & \textbf{Topology / Span} & \textbf{KS (k=15)} & \textbf{KS (k=37)} & \textbf{KS (k=68)} & \textbf{KS (k=100)} \\
\midrule
AsymBlip & Ultra-Short ($\sim 15$ patches) & 0.000 ($p=1.00$) & 0.098 ($p=0.66$) & \textbf{0.117} ($p=0.56$) & 0.043 ($p=0.90$) \\
SpiralBurst & Mid-Band ($\sim 74$ patches) & 0.807 ($p<0.01$) & 0.927 ($p<0.01$) & \textbf{0.963} ($p<0.01$) & 0.935 ($p<0.01$) \\
HarmonicComb & Broadband ($>300$ patches) & 0.987 ($p<0.01$) & 0.976 ($p<0.01$) & \textbf{0.993} ($p<0.01$) & 0.983 ($p<0.01$) \\
\bottomrule
\end{tabular}
\end{table*}

The empirical KS matrix provides immediate, quantifiable evidence of the Patch-Level resolution capabilities and its physical limits:

\begin{enumerate}
    \item \textbf{The Spatial Diffraction Limit (AsymBlip):} As anticipated by the physical constraints of the $14 \times 14$ ViT convolutions, ultra-short transients like the \textit{AsymBlip} fail to significantly deviate from the noise GEV distribution across \textit{any} $k$ configuration. Even at the highest amplitude ($\text{SNR} \approx 403$), the KS test fails to reject the null hypothesis, yielding $p$-values consistently above the significance threshold ($p \in [0.569, 1.000]$). The signal footprint is simply too small to dominate even a single patch, resulting in unavoidable intra-patch dilution and confirming mathematical blindness to sub-patch events.
    \item \textbf{Topological Mapping (SpiralBurst):} For mid-band transient signals, the KS statistic exhibits a clear maximization curve. The separability peaks precisely at $k=68$ ($\text{KS}=0.963$), before degrading at $k=100$ ($\text{KS}=0.935$). This empirical drop-off strictly validates that increasing $k$ beyond the signal's topological area re-introduces the Signal Dilution Effect by incorporating surrounding null-noise patches into the pooling operation.
    \item \textbf{Broadband Saturation (HarmonicComb):} Contrary to concerns that spectral lines might be entirely suppressed by the whitening high-pass filter, the \textit{HarmonicComb} demonstrates extreme separability ($\text{KS} > 0.97$) across the entire $k$-sweep. While \citet{cirfeta2026b} reported $\text{Recall}=0.00$ for \textit{HarmonicComb} under \texttt{[CLS]}-based global scoring, patch-level Top-$k$ scoring yields $\text{KS}=0.993$ for the same morphology. Because its morphological footprint is massive (spanning horizontally across the entire spectrogram), it provides a sufficient number of highly anomalous patches to saturate the novelty metric even at low $k$ pooling, slightly peaking at $k=68$.
\end{enumerate}

\begin{figure}[ht]
    \centering
    \includegraphics[width=\linewidth]{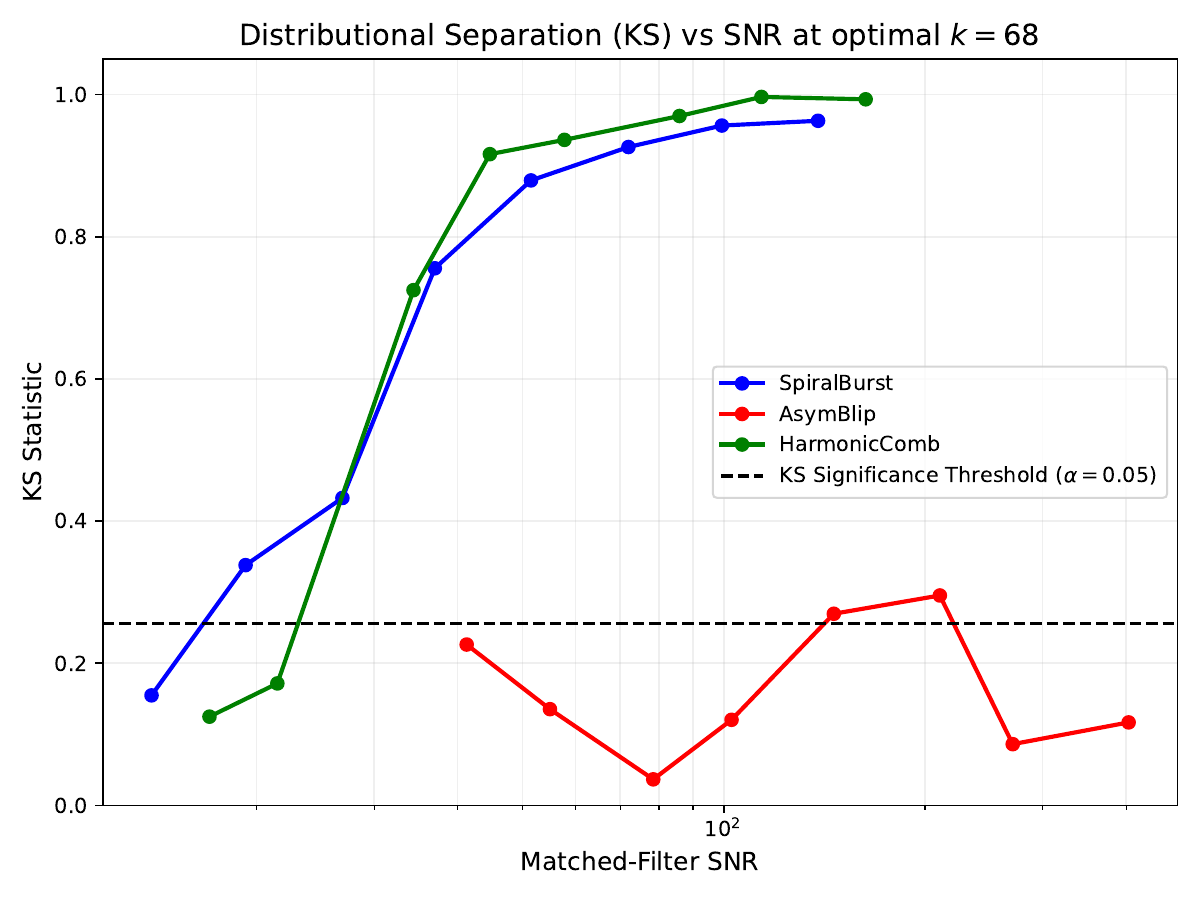}
    \caption{Kolmogorov-Smirnov (KS) statistic vs. Matched-Filter SNR for \textit{AsymBlip}, \textit{SpiralBurst}, and \textit{HarmonicComb} at optimal $k=68$. The dashed line indicates the threshold for statistical significance ($\alpha=0.05$). The \textit{SpiralBurst} experiences a transition to high separation at $\text{SNR} \approx 37$, whereas \textit{AsymBlip} remains strictly non-significant across the entire domain, mathematically confirming the ViT spatial diffraction limit.}
    \label{fig:ks_curve}
\end{figure}

Due to the severity of the Generalized Extreme Value (GEV) threshold required to maintain an $\text{FPR} < 1\%$ in the presence of extreme heavy-tailed non-Gaussian noise, the Boolean detection metric (Recall) is unstable and physically limited. Therefore, we evaluate the architectural sensitivity exclusively via distributional separation (Kolmogorov-Smirnov statistic) \citep{smirnov1948, kolmogorov1933}.

\subsection{Saliency Map Validation}
We validated the purely spatial Topological Saliency Map on isolated samples.

\begin{table}[ht]
\centering
\caption{Saliency validation metrics on isolated samples.}
\label{tab:saliency}
\begin{tabular}{lccc}
\toprule
\textbf{Sample} & \textbf{Hotspot} & \textbf{Max/Mean} & \textbf{Test Outcome} \\
\midrule
Scattered\_Light & (22, 11) & 8.54 & PASS \\
SpiralBurst (SNR 135) & (17, 23) & 5.11 & PASS \\
No\_Glitch (Noise) & (8, 9) & 5.54 & N/A (Visualizer) \\
\bottomrule
\end{tabular}
\end{table}

The decoupling strategy proved mathematically robust: edge artifacts were entirely ignored. The \texttt{Scattered\_Light} hotspot mapped perfectly to $\text{row}=22$, corresponding to the low-frequency scattering arch physically expected in the 50-200 Hz band. The injected \textit{SpiralBurst} was correctly located at the center $(17, 23)$.

\begin{figure}[ht]
    \centering
    \includegraphics[width=\linewidth]{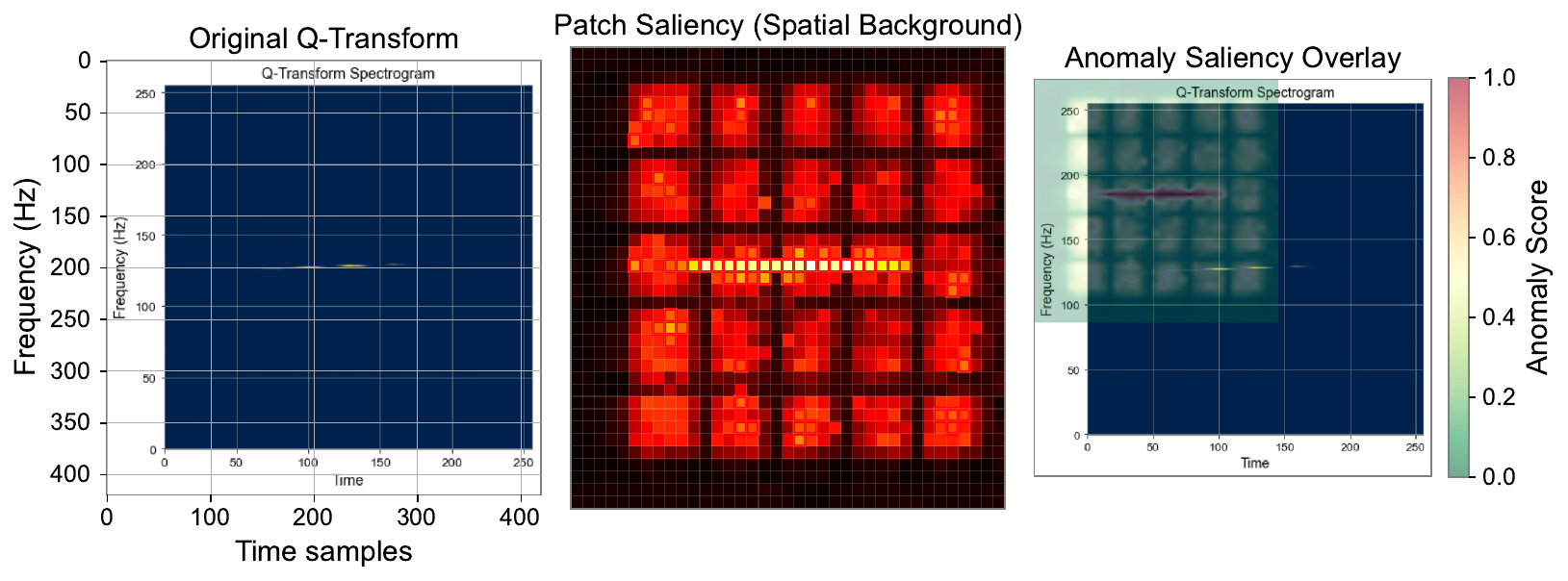}
    \caption{Topological Saliency Map applied to an injected \textit{SpiralBurst} ($\text{SNR} \approx 138$). The spatial mapping isolates the morphological footprint of the transient, entirely ignoring Q-Transform boundary artifacts via purely spatial distance evaluations against the null median matrix.}
    \label{fig:saliency}
\end{figure}

Notably, the \texttt{No\_Glitch} background noise sample yielded a Max/Mean ratio of 5.54, slightly higher than the massive \textit{SpiralBurst} (5.11).

\section{Discussion} \label{sec:discussion}

\subsection{Mitigating the Signal Dilution Barrier}
The results of the Micro-MDC confirm that extracting features at the $14\times14$ patch level mitigates the Signal Dilution Effect for transient durations $>0.5$s. The Kolmogorov-Smirnov test proves that the embedding shift is highly significant ($p \ll 0.05$) exactly where the \texttt{[CLS]} token was statistically blind. The $k$-sweep directly correlates the optimal detection statistic with the physical area of the anomaly, transitioning $k$ from an arbitrary hyperparameter to a physically motivated mapping operator.

\subsection{The Nature of DINOv2 Embedding Space}
The analysis of the Max/Mean ratio (Table \ref{tab:saliency}) yields a critical insight into the latent geometry of Vision Transformers. The observation that pure strain background noise produces localized similarity spikes ($\times 5.54$) comparable to extreme ($\text{SNR}=135$) injections proves that the cosine similarity hypersphere of DINOv2 is severely non-isotropic. In high-dimensional embedding spaces, extreme order statistics (the tails of the background distribution) manifest as highly localized variance in individual patch tokens. Coherent morphological signals, conversely, exhibit correlated spatial variance across contiguous patches. The comparable Max/Mean ratios between \texttt{No\_Glitch} (5.54) and \textit{SpiralBurst} (5.11) indicate that the DINOv2 patch embedding space does not preferentially concentrate anomaly energy for physical signals versus stochastic noise fluctuations at the single-frame level. This confirms that the saliency map functions as a topological visualizer, not a binary classifier.
Consequently, Topological Saliency Maps must never be utilized as single-frame binary detectors; their purpose is strictly limited to post-hoc morphological visualization of frames already validated by the global VQ-index GEV trigger.

\subsection{Limitations and Roadmap}
We acknowledge three primary limitations:
\begin{enumerate}
    \item \textbf{Patch-Size Temporal Resolution Limit}: Transients vastly shorter than the ViT patch size (e.g., \textit{AsymBlip}, $<0.1$s) remain statistically unresolved ($\text{KS} \le 0.117$) regardless of patch-level scoring. They exist beneath the spatial resolution limit of the $37\times37$ grid.
    \item While the Micro-MDC maps the optimal separation to a specific $k$ (e.g., $k=68$ for \textit{SpiralBurst}), a production pipeline will require adaptive $k$-sweeping or an ensemble aggregation approach to simultaneously capture highly variable morphological footprints without incurring $k$-miss-match dilution.
    \item The strain-domain whitening barrier suppresses morphological continuity even before the transformer encoding phase. Multi-scale temporal windowing is necessary for full pipeline integration.
\end{enumerate}

In Phase 4 (Production Scan), we will deploy \textbf{Multiple Instance Learning (MIL) via Mean Pooling}. The Top-$k$ anomalous patches identified by the VQ index will be aggregated into a single 384-dimensional representation. This avoids the dimensional collapse of DPMM clustering while retaining the undiluted anomaly signal of the transient.

\section{Conclusion} \label{sec:conclusion}

The application of frozen Vision Transformers to gravitational-wave time series requires careful treatment of non-stationary spatial geometries. Global average pooling natively implemented by the \texttt{[CLS]} token induces a Signal Dilution Effect that effectively blinds the model to short transients.
By pivoting to a Vector-Quantized Reference Index and implementing a Top-$k$ patch-level scoring algorithm, we have demonstrated a statistically robust resolution to this barrier. Patch-level scoring produces statistically significant separation ($\text{KS}=0.963$, $p \ll 0.05$ for optimal $k$) where \texttt{[CLS]} global pooling was completely blind, demonstrating partial resolution of the signal dilution effect at the architectural level. This framework provides the topological extraction necessary for DPMM to discover unmodeled transient populations in LIGO O4a.

\vspace{4mm}
\noindent\textbf{Data and Code Availability} \\
The strain data used in this study are publicly available via the Gravitational Wave Open Science Center (GWOSC). The complete source code, vector-quantized index matrices, and reproduction scripts for the Micro-MDC are available on GitHub: \url{https://github.com/lucacirfeta/dante-gravi-signal-ml} (DOI: \href{https://doi.org/10.5281/zenodo.20121860}{10.5281/zenodo.20121860}).

\vspace{4mm}
\noindent\textbf{Acknowledgments} \\
This research has made use of data or software obtained from the Gravitational Wave Open Science Center (\url{gwosc.org}), a service of LIGO Laboratory, the LIGO Scientific Collaboration, the Virgo Collaboration, and KAGRA.

\end{document}